# Configuration Defects in Kubernetes

Yue Zhang, Uchswas Paul, Marcelo d'Amorim, and Akond Rahman, *Member, IEEE*

**Abstract**—Kubernetes is a tool that facilitates rapid deployment of software. Unfortunately, configuring Kubernetes is prone to errors. Configuration defects are not uncommon and can result in serious consequences. This paper reports an empirical study about configuration defects in Kubernetes with the goal of helping practitioners detect and prevent these defects. We study 719 defects that we extract from 2,260 Kubernetes configuration scripts using open source repositories. Using qualitative analysis, we identify 15 categories of defects. We find 8 publicly available static analysis tools to be capable of detecting 8 of the 15 defect categories. We find that the highest precision and recall of those tools are for defects related to data fields. We develop a linter to detect two categories of defects that cause serious consequences, which none of the studied tools are able to detect. Our linter revealed 26 previously-unknown defects that have been confirmed by practitioners, 19 of which have already been fixed. We conclude our paper by providing recommendations on how defect detection and repair techniques can be used for Kubernetes configuration scripts. The datasets and source code used for the paper are publicly available online.

**Index Terms**—configuration, container orchestration, defect, devops, Kubernetes

———————————— ◆ ————————————

## 1 INTRODUCTION

THE use of multiple containers to deploy software projects is a common practice today [83], e.g., Paypal uses 200,000 containers to speed up financial transactions [58]. Setting up and managing multiple containers manually is considered impractical and prone to errors [16], [65], [74]. For that reason, the practice of container orchestration advocates for automated management of containers with tools, such as Kubernetes [64] that has yielded benefits for organizations [46]. OpenAI reported that Kubernetes enabled a reduction of deployment time from "a couple of months" to "two or three days" [46]. Kubernetes usage aided Adidas to reduce the load time for their e–commerce website by half, and increase the release frequency from once every 4~6 weeks to 3~4 times a day [46].

Unfortunately, Kubernetes configuration scripts are not immune to defects. In March 2023, the social media platform Reddit experienced a 5 hour–long outage that impacted millions of its users [41], [70]. The outage occurred because of a defect in a configuration script affecting the network traffic between containers [40], [41]. Figure 1 presents an YAML code snippet showcasing how certain Kubernetes–related configurations were specified when the outage occurred. The defect is due to the incorrect definition of configuration options `nodeSelector` and `peerSelector`, which used the value `node-role.kubernetes.io/master` instead of `node-role.kubernetes.io/control-plane`. The string `master` in the configuration value became obsolete with the release of Kubernetes 1.24 [45]. The entities `nodeSelector` and `peerSelector` are responsible to route the network traffic across containers. As a result of this defect, traffic was routed to a destination that does

```
metadata:
    annotations:
...
spec:
    asNumber: 0
    nodeSelector: has (node-role.kubernetes.io/master)
    nodeSelector: has (node-role.kubernetes.io/control-plane)
    peerIP: ' '
    peerSelector: has (node-role.kubernetes.io/master)
    peerSelector: has (node-role.kubernetes.io/control-plane)
```

Fig. 1: Excerpt of the configuration defect that caused the Reddit outage [41].

not exist, resulting in the outage. This defect illustrates the importance of understanding configuration defects in Kubernetes–related computing infrastructure.

Our paper presents an empirical study about configuration defects in Kubernetes with the goals of assisting practitioners in preventing defects and guiding researchers in developing automated tools to detect those defects. The results of the study enable researchers and practitioners (i) to gain insights about the defects in Kubernetes–based computing infrastructure; (ii) to assess the capabilities of existing tools in identifying defects; and (iii) to develop techniques to identify latent defects that occur during Kubernetes–based configuration management. While the importance of defect categorization have been well–acknowledged in software engineering research [15], [35], [69], a systematic characterization of defects related to Kubernetes configuration management remains under explored.

We answer the following research questions:

- **RQ1 [Categories]**: What are the categories of defects in Kubernetes configuration management?
- **RQ2 [Consequences and Fix Patterns]**: What categories of consequences and fix patterns map to defects that occur during Kubernetes configuration management?
- **RQ3 [Tool Support]**: How frequently do static analysis tools support the detection of defects that occur during Kubernetes configuration management?

———————————————
Yue Zhang is with Graduate School of Computer Science and Software Engineering, Auburn University, Auburn, Alabama, USA
Uchswas Paul is with Graduate School of Computer Science, NC State University, Raleigh, NC, USA
Marcelo d'Amorim is with the Faculty of Computer Science, NC State University, Raleigh, NC, USA
Akond Rahman is with the Faculty of Computer Science and Software Engineering, Auburn University, Auburn, Alabama, USA

We analyze 719 defects that occur in 2,260 configuration scripts mined from 185 open source software (OSS) repositories. We use a qualitative analysis technique called open coding [68] with the obtained data to derive defect categories, consequences, and fix patterns. Using the data, we systematically evaluate the defect detection capabilities of 8 publicly available static analysis tools for Kubernetes. Our empirical study provides insights on the nature of configuration defects and identify opportunities for developing defect detection techniques for Kubernetes. For example, we find that 533 of the 719 defects to cause crashes, incorrect operations, or outages. We construct a linter that detects two categories of defects that cause serious consequences, such as crashes and outages. These two defect categories are not detected by any of the 8 studied tools. With the help of the linter, we have identified 26 previously–unknown defects that have been confirmed by practitioners, and 19 have already been fixed.

**Contributions**: We list our contributions as follows:

- An evaluation on the performance of static analysis tools to detect defects that occur during Kubernetes configuration management (Section 5.1.2);
- A categorization of consequences and fix patterns for defects that occur during Kubernetes configuration management (Section 4.2); and
- A list of derived defect categories for Kubernetes configuration management (Section 3.2).

**Dataset Availability**: Datasets and source code used in our paper are publicly available online [91]. The dataset contains data where each of the 719 defects are mapped to their corresponding defect category, consequence, and fix pattern. Source code to construct our linter is available. The defect reports that we submitted are also shared anonymously.

## 2 BACKGROUND

Kubernetes is the most popular tool to implement container orchestration. Any computing infrastructure managed by Kubernetes is referred to as a Kubernetes cluster [47]. Kubernetes uses objects to provision the cluster computing infrastructure. An object is a persistent entity representing the state of the cluster. A pod is a common kind of object; it is the most fundamental deployment unit that groups multiple containers together. Configurations for pods and other Kubernetes entities are specified using configuration scripts that are typically written in the YAML format. As Figure 2 shows, the API server stores configurations in a database called 'etcd'. With the provided configurations, the API server decides which pods can host the given containers. A controller and scheduler are automated agents that control the state of the Kubernetes to identify a suitable node for a pod. A configuration script can either be a Kind script or a Helm script [64].

**Kind script:** Kind scripts contains configurations for kind, which is a specific type of Kubernetes object. Kind scripts are executed using Kubernetes–provided utilities, such as 'kubectl' [47]. Listing 1 shows an example of a pod specified with a Kind script. This script defines a pod that runs a single container using the image 'myimage'.

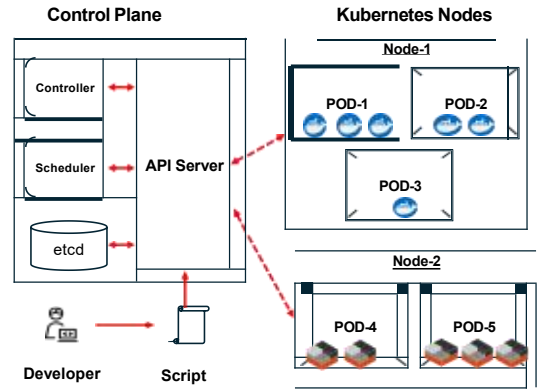

Fig. 2: An overview of the components in a Kubernetes cluster.

```
1   apiVersion: v1
2   kind: Pod
3   metadata:
4     name: mypod
5   spec:
6     containers:
7     - name: mycontainer
8       image: myimage
9       ports:
10        - containerPort: 80
```

Listing 1: An example of a Kind script.

```
1   # configuration values defined in a Helm script, [values.yaml]
2   replicaCount: 2   --> Configuration value for spec.replicas
3   service:
4     portName: https   --> Configuration value for spec.ports.name
5     portHttps: 80   --> Configuration value for spec.ports.port
6   # configuration values used by a Helm template
7   spec:
8     replicas: {{ .Values.replicaCount }}   --> Template directive
9     ports:
10      - name: {{ .Values.service.portName }}
11      - port: {{ .Values.service.portHttps }}
12      - protocol: TCP
```

Listing 2: An example of a Helm script.

**Helm script:** Helm is a package manager for Kubernetes that simplifies configuration management for Kubernetes [10]. A Helm script is developed using YAML, and a group of Helm scripts is referred to as a Helm chart. In a Helm chart, variables and default configuration values are defined in a script labeled as 'values.yaml' [10]. These variables and configuration values are loaded dynamically into scripts called 'templates' through template directives [10]. Listing 2 shows an example of a template.

## 3 RQ1: CATEGORIES OF DEFECTS IN KUBERNETES CONFIGURATION SCRIPTS

We provide the methodology and results for RQ1 respectively, in Sections 3.1 and 3.2.

### 3.1 Methodology

We use the following steps:

#### 3.1.1 Identify Defects from OSS projects

We follow three steps to identify defects.

**Step#1 - Mine OSS repositories from GitHub**: We identify defects by mining OSS repositories hosted on GitHub, which is the most popular code hosting platform [59]. We mine repositories using the GHTorrent archive [30] that is hosted on Google Big Query. However, as publicly–available GitHub repositories are susceptible to quality issues [59], we apply the following filtering criteria: Criterion-1: Repos–itory must be publicly available and contain the 'Kuber–netes' label to ensure that the repositories are Kubernetes relevant [64]; Criterion-2: At least 10% of the files in the repository are YAML files and each file must use Kuber–netes objects (e.g., Pod, Service, Deployment, etc) to collect repositories that contain sufficient amount of configuration scripts for analysis; Criterion-3: The repository is not a copy of another repository; and Criterion-4: The repository has at least ten contributors. We use a threshold of ten contributors to ensure a higher likelihood that the repository represents a more collaborative and active project, reducing the chances of including repositories used for academic or personal projects. Prior research [63] has also used the threshold of at least 10 contributors.

As shown in Table 1, we collect 185 OSS Kubernetes repositories from GitHub repositories. We clone the master branches of the 185 repositories. We provide attributes of the mined 185 repositories in Table 2. In all we collect 44,401 configuration scripts.

**Step#2 - Mine Commits and Issue Reports from 185 OSS Kubernetes repositories**: We download the 185 OSS Kubernetes repositories on March 2024 to conduct our anal–ysis. From the downloaded repositories, we mine 417,598 commits and 140,872 issue reports. To identify commits and issue reports that are related to defects, we use the following steps: Step-1: We filter issue reports by checking if the issue is closed and have a pull request to ensure we have sufficient content to derive fix patterns; Step-2: We apply a keyword search similar to prior work [63]. We use following keywords: 'bug', 'defect', 'error', 'fault', 'fix', 'flaw', 'incorrect', 'issue', and 'mistake' to ensure commits and issue reports are related to a defect; Step-3: We inspect the files modified in each commit and issue report to ensure commits and issue reports are related to Kubernetes config–uration management; and Step-4: We exclude commits that are duplicates of others. In all, we identify 66 commits and 1,941 issue reports that include defect–related keywords.

**Step#3 - Detect Defects by Applying Qualitative Analysis**: We conduct qualitative analysis to identify defects from defect–related commits and issue reports. The rationale is that relying solely on keyword search can result in false pos–itives. To identify defects, we use the IEEE definition [37]: "an imperfection or deficiency in the code that needs to be repaired".

<u>Criteria to Identify Defects</u> – For defect identification, the rater applies the following criteria: (i) problematic code exists in the commit message or the issue report; (ii) prob–lematic code leads to an incorrect or undesired consequence that is explicitly expressed by a practitioner; (iii) the commit message or issue content describes an immediate conse–quence of the defect; and (iv) the problematic code was repaired. By applying this criteria, we identify that 52 of the 66 commits and 681 of the 1,941 issue reports to be related with defects.

<u>Criteria to Identify Configuration Defects</u> – The rater in–spects if any of the following criterion is satisfied: (i) the defect resides in a configuration script; (ii) the defect occurs when provisioning Kubernetes resources, or managing Ku–bernetes resources, or monitoring Kubernetes resources; and (iii) the defect is related to a Kubernetes configuration. Us–ing this criteria, we identify 719 defects. Of these 719 defects 52 and 667 are respectively, obtained from 52 commits and 681 issue reports. Figure 3 shows the distribution of defects across 185 repositories. We observe 77.8% of the studied repositories to include =< 5 defects.

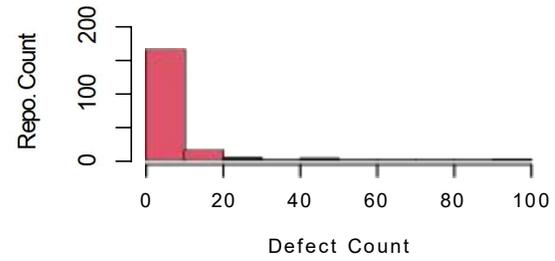

Fig. 3: Distribution of defects.

### 3.1.2 Derive Defect Categories

We employ a qualitative analysis method known as open coding [68] to derive defect categories. Open coding in–volves recognizing patterns in unstructured text to establish categories [68]. The first and second author individually applies open coding with 52 defects from 52 defect–related commits and 667 defects from 667 defect–related issue re–ports. While applying open coding, each rater applies or–thogonality, i.e., derive the categories so that do not overlap. Each rater examines messages and code changes for each commit, as well as title, description, comment, pull request, and code changes for each issue report.

Each rater creates a category with a short definition, which we use to identify and resolve differences in labeling. Dis–agreements occurred as the first and last author respectively, identifies 20 and 12 categories, where 11 are identified by both, 9 identified by the first author, and 1 identified by the last author. Amongst the 10 disagreements, 3 occurred because of naming issues, e.g., 'conditional operator' and 'conditionals', and 7 occurred because of definition overlap, e.g., 'access control' is a sub–category of 'security'. For disagreements, the last author's decision is final. The raters disagree on 7 categories. The Cohen's Kappa [17] is 0.67, suggesting 'substantial' agreement [51]. Disagreements are resolved by the last author and their decision is final.

### 3.2 Answer to RQ1

We answer RQ1 by reporting the defect categories (Sec–tion 3.2.1), their frequency (Section 3.2.2).

### 3.2.1 Answer to RQ1: Defect Categories

We identify 15 defect categories, which we characterize with examples obtained from our OSS repositories.

**I Conditional**: This kind of defect manifests when develop–ers use incorrect operators or operands in conditional state–ments, such as if–else blocks. Listing 3 shows an example defect [26] where an improper operand, i.e., 'and' is used in

TABLE 1: Filtering of OSS Repositories

| | |
|---|---|
| Initial Repository Count | 14,747,836 |
| Criterion–1 (Available and relevant) | 1,410 |
| Criterion–2 (>= 10% Configuration scripts) | 1,087 |
| Criterion–3 (Not a copy) | 1,079 |
| Criterion–4 (Contributors >=10) | 185 |
| Final Repository Count | 185 |

TABLE 2: Dataset Attributes

| Category | Data |
|---|---|
| Total Repositories | 185 |
| Total Commits | 417,598 |
| Total Developers | 21,559 |
| Kind Scripts | 37,147 |
| Helm Scripts | 7,254 |
| Total Kubernetes Scripts | 44,401 |
| Total Size (LOC) | 51,282,124 |
| Total Count of Issue Reports | 140,872 |
| Total Count of Stars | 398,347 |
| Time Span | 06/2014 – 03/2024 |

```
{{- if and .Values.certificates.autoGenerated ( not
    .Values.certificates.certManager.enabled ) }}
{{- if or (and .Values.certificates.autoGenerated (not
    .Values.certificates.certManager.enabled))
    (.Values.permissions.operator.restrict_secret) }}
apiVersion: rbac.authorization.k8s.io/v1
```

Listing 3: Example of a conditional–related defect.

```
command:
-   - longhornio/backing-image-manager: v2_20210820_patch2
+   - longhornio/backing-image-manager:v2__20221027
```

Listing 4: Example of a CLA–related defect.

```
resources:
  limits:
-    memory: 256Mi
+    memory: 550Mi
```

Listing 5: Example of a resource–related defect.

an if–else block. Due to this defect, the application fails to startup.

II **Container Provisioning**: This defect category occurs when developers provision containers for pods. There are two sub–categories: (i) Command Line Arguments (CLA): This defect category occurs due to specifying erroneous command line arguments. Arguments can be provided either from command line or using the command or args property. Listing 4 shows a CLA–related defect [55] where an erroneous argument is provided for command. Due to this defect, the image is unable to be recovered after being deleted. (ii) Resources: This defect occurs because of provisioning resources that are unspecified, under–specified, or over–specified. Listing 5 shows a resources–related defect [31] where resource limits are under–specified, i.e., 256 Mebibytes (Mi) is used instead of 550Mi. This defect causes a hang.

```
kind: ClusterServiceVersion
...
  - name: OPERATOR_NAME
-   image: quay.io/jaegertracing/jaeger-operator:v1.29.0
+   image: quay.io/jaegertracing/jaeger-operator:1.29.0
```

Listing 6: Example of a CR–related defect.

III **Custom Resource**: This defect category occurs when developers incorrectly manage custom resources (CRs) in Kubernetes. CRs are extensions of the Kubernetes API that allow developers to create and manage new kinds of resources beyond what Kubernetes offers by default [47]. Listing 6 shows an example defect [39] where an incorrect image tag is configured for the ClusterServiceVersion CR.

IV **Data Fields**: This defect category occurs when the data fields are improperly handled in scripts. We identify five sub–categories: (i) Base64 String and Encoding (BSE): This defect occurs due to the misuse of Base64 encoding. Figure 4a shows a BSE–related defect [9] because of not using Base64 encoding with 'b64enc'. (ii) Incorrect Data Types (IDT): This defect occurs because of using incorrect data types. Figure 4b shows an IDT–related defect [23] that causes a hang. The defect occurred because of missing quote, which causes annotations to be interpreted as numbers instead of strings. (iii) Incorrect URL Path Types (IUPT): This defect occurs due to misuse of pathType, an attribute used to route incoming traffic to the back–end services. Figure 4c shows an IUPT–related defect [88] which results in a dashboard failing to load. (iv) Syntax: This defect occurs due to syntax errors. Figure 4d shows a syntax–related defect [44] where incorrect indentation is used. (v) Violation of Restrictions (VR): This defect occurs due to the failure to adhere to the specific technical rules and constraints enforced by Kubernetes on resource definitions and configurations. These restrictions include, but are not limited to, name length, allowed characters, and the correct format of values. Figure 4e shows a VR–related defect [33] where a dynamically generated name exceeds the maximum length of 63 characters.

V **Entity Referencing**: This defect category occurs when Kubernetes entities, such as names

```
metadata:
  labels:
-   control-plane: controller-manager
+   control-plane: argocd-operator
```

Listing 7: A defect related to entity referencing.

and labels are incorrectly referenced or entities that are referred to do not exist. Listing 7 shows an example defect [8] where the incorrect label 'controller–manager' is provided instead of 'argocd–operator'.

```
containers:
  - name: main
-   image: docker.io/aquasec/trivy:0.34.0
+   image: {{ .Values.trivy.repository }}:{{ .Values.trivy.tag }}
```

Listing 8: A defect related to incorrect Helming.

VI **Incorrect Helming**: This defect category occurs when users hard–code configuration values in templates. Hard–coding configuration values in templates is considered as an anti–pattern in Kubernetes [12]. Listing 8 shows an example defect [4], where the configuration value is hard–coded in a template. Due to this defect, the user–provided image value is never applied.

```
subjects:
-   namespace: {{ .Release.Namespace }}
+   namespace: {{ include "opentelemetry-collector.namespace" . }}
```

Listing 9: Example of a namespace–related defect.

VII **Namespaces**: This defect category occurs when an incorrect namespace is used. Namespaces provide a mechanism for isolating groups of resources within a single Kubernetes cluster by separating different environments [47]. If resources are placed in a namespace that is different from the objects that use them, then the referencing objects will not be able to access these resources, leading to application failures. Listing 9 shows an example defect where the namespace is incorrect due to an incorrect template directive. As a result, the deployed application of interest results in a crash.

```
-    kind: ClusterRole
-    name: argocd-server
-    ...
-    kind: ClusterRoleBinding
-    name: argocd-server
-    ...
     kustomization.yaml:
     resources:
-      - argocd-server-clusterrole.yaml
-      - argocd-server-clusterrolebinding.yaml
+      - ./application-controller
```

Listing 10: A defect related to orphanism.

**VIII Orphanism**: This defect category occurs when either resources in a pod are not properly de–allocated, or when resources are deployed but not referenced by any other resources. Listing 10 shows an example defect [7] where a ClusterRoleBinding object references a non–existent service account, i.e., 'argocd–server'. This defect leads to a resource leak, as the orphaned ClusterRole and ClusterRoleBinding continue to consume cluster resources unnecessarily.

**IX Pod Scheduling**: This defect category occurs when developers incorrectly use pod scheduling mechanisms, such as affinity. Affinity is a set of rules that assign pods to nodes based on certain criteria, such as node labels or the location

```
-    affinity: {}
+    affinity:
+      nodeAffinity:
```

Listing 11: A defect related to pod scheduling.

of other pods [47]. Listing 11 shows an example defect [49] where affinity is missing. This causes a pod to be unexpectedly scheduled on the 'fargate' node, leading to resource contention between pods.

**X Probing**: This defect category that occurs when probing is incorrectly handled in scripts. Kubernetes provides two health check probes namely, liveness probes and readiness probes to monitor the health status of the

```
-    tcpSocket:
+    livenessProbe:
+      httpGet:
+        path: /healthz
```

Listing 12: Example of a probing–related defect.

provisioned containers [47]. Listing 12 shows an example defect [6] where the configurations for a liveness probe is missing. Due to this defect, the pod could not automatically recover from an error status when a failure occurred, leading to an outage.

**XI Property Annotation**: This defect category occurs when developers use user–defined

```
-    "helm.sh/hook": pre-install
+    "helm.sh/hook": pre-install, pre-upgrade
```

Listing 13: A defect related to property annotation.

annotations incorrectly. Annotations are used to attach arbitrary non–identifying metadata to objects [47]. Unlike labels, which are used to organize and select subsets of objects, annotations are not used to identify and select objects. Instead, they are used to store additional information that may be used by external libraries. Listing 13 shows an example defect [27] because of incorrectly using the pre-upgrade annotation.

**XII Security**: This category includes defects that violate the principle of confidentiality, integrity, or availability. The four sub–categories are: (i) Access Control (AC): Access control is defined as the technique that regulates who or what can view or use resources in a computing environment. If access control is improperly configured, it can lead to creation of over–privileged and under–privileged entities. Over–privileged entities, such as users or processes may perform unauthorized actions, access sensitive data, or disrupt the operation of the system [24]. Under–privileged entities can lead to availability issues, such as not being able to access needed cluster data. Figure 5a shows an AC–related defect [50] that occurred because of using '*' that allows unauthorized users to gain access to sensitive data. (ii) Privileged Ports (PP): This defect occurs due to the use of a privileged port number. Using privileged ports that are typically below 1024 requires higher privileges, which can increase security risks, such as privilege escalation, if not properly managed [5]. Privilege escalation can expose the system to attacks, as they may require running applications or containers with more access than necessary allowing a malicious user to gain unauthorized control [62]. Figure 5b shows a PP–related defect [48] where a privileged port number 433 is used. (iii) Exposure of Sensitive Data (ESD): This defect occurs due to exposure of sensitive data in scripts. Figure 5c shows an ESD–related defect [61] where a plain string can be mistakenly passed to the Secret entity. (iv) Security Context (SC): This defect that occurs due to privileged securityContext or a missing securityContext. A securityContext is a Kubernetes entity that determines the user IDs, group IDs, and whether the container runs as a privileged user. An improperly configured securityContext can result in containers running with unnecessary privileges, increasing the risk of privilege escalation, unauthorized access, and potential compromise of the system [54]. Figure 5d shows a SC–related defect [53] where 'runAsUser' is missing for securityContext, which causes the container running with root privileges. Due to this defect, malicious users could gain unauthorized access.

**XIII Unsatisfied Dependency**: This defect category occurs when execution of scripts are dependent on one or multiple prerequisites, such as network–related dependencies and container im-

```
     accessModes:
-      - ReadWriteOnce
+      - {{ .Values.accessMode }}
     values.yaml:
+    accessMode: ReadWriteMany
```

Listing 14: A defect related to unsatisfied dependency.

ages. Listing 14 shows an example defect [38] where scaling up pods on different nodes fails due to the missing precondition ReadWriteMany. The ReadWriteMany access mode in the persistent volume claim configuration allows multiple nodes to read and write simultaneously, which is a crucial precondition for scaling up pods across different nodes.

```
stringData :
  username : "{{ .vsphereUsername }}"
  password : "{{ .vspherePassword }}"
data :
  username : {{ .vsphereUsername | b64enc }}
  password : {{ .vspherePassword | b64enc }}
```
a

```
node : {{ $sts .node }}
node : {{ $sts .node | quote }}
```
b

```
path : {{ . }}
pathType : ImplementationSpecific
pathType : Prefix
```
c

```
{{ - toYaml .Values .volumes .keda .
   extraVolumeMounts | nindent 12 }}
{{ - toYaml .Values .volumes .keda .
   extraVolumeMounts | nindent 10 }}
```
d

```
volumeMounts:
  - name : data -{{ .Release .Namespace }}
  - name : data -{{ .Release .Namespace |
    trunc 58 | trimSuffix "-" }}
```
e

Fig. 4: Examples of defects related to data fields. Figures 4a, 4b, 4c, 4d, and 4e respectively, presents examples of defects related to BSE, IDT, IUPT, syntax, and VR.

```
rules:
  apiGroups:
-   - "*"
+   - ""
+   - events .k8s .io
```
a

```
args:
   --cert -dir=/tmp
-  --secure -port=443
+  --secure -port=4443
```
b

```
data:
{{ - if eq (typeOf .Values .alertmanager .config) "string" }}
{{ - if .Values .alertmanager .stringConfig }}
  alertmanager .yaml : {{ tpl (.Values .alertmanager .stringConfig)
    . | b64enc | quote }}
{{ - else if eq (typeOf .Values .alertmanager .config) "string" }}
  alertmanager .yaml : {{ tpl (.Values .alertmanager .config)
    . | b64enc | quote }}
```
c

```
securityContext:
  runAsNonRoot : true
  runAsUser : 65534
```
d

Fig. 5: Examples of security defects. Figures 5a, 5c, 5b, and 5d respectively, presents examples of defects related to AC, PP, ESD, and SC.

**XIV Version Incompatibility**: This defect category occurs when developers use APIs or Kubernetes objects that are no longer supported by Kubernetes and its API. Listing 15 shows an example defect [1] where a deprecated API version 'extensions/v1beta1' is used. Due to this defect, the configuration script fails to be executed and leads to a crash.

```
- apiVersion: extensions/v1beta1}
+ apiVersion: apps/v1
  kind: Deployment
```
Listing 15: A defect related to version incompatibility.

**XV Volume Mounting**: This defect category occurs when developers incorrectly mount storage for applications that are managed by Kubernetes. Listing 16 shows an example defect [2] where the 'zk–data' volume is incorrectly mounted instead of 'zk–datalog'. This defect resulted in a crash.

```
  - mountPath: /datalog
-   name: zk-data
+   name: zk-datalog
```
Listing 16: A defect related to volume mounting.

Of the identified 15 defect categories, 6 are unique to Kubernetes: custom resource, namespaces, orphanism, pod scheduling, property annotation, and volume mounting. These categories are unique because they focus solely on defects that occur in YAML–based configuration files, which do not overlap with prior work that focuses on defects in Kubernetes operators. Unlike, operators that are custom controllers [47], configuration scripts are used to configure built–in Kubernetes entities, such as pods, custom resources, and namespaces. Custom resources in Kubernetes are used to extend the Kubernetes API by allowing developers to define and manage their own custom resource types through Custom Resource Definitions (CRDs). Namespace in Kubernetes is used to isolating resources by providing logical separation within a cluster. Orphanism is unique to pods, where resources of pods, such as CPU and memory are left unused or unlinked due to improper cleanup or misconfiguration. Pod scheduling is unique to Kubernetes pods, which is conducted when the scheduler assigns pods to appropriate nodes based on resource availability, constraints, and policies. Property annotation is performed for Kubernetes resources to provide metadata or configuration details, such as specifying labels, or custom behaviors. Mounting volume is applicable for Kubernetes pods, where the volume tag is used to define the storage volumes and attach them to containers.

#### 3.2.2 Results for RQ1: Frequency

We present the count of defects for each defect category in Table 3, organized alphabetically by category names. The most frequently occurring category is entity referencing. 'N/A' denotes categories without sub–categories. 'Category Total' represents the overall count of defects for categories with sub–categories.

## 4 RQ2: Consequences and Fix Patterns

We provide the methodology and results for RQ2 respectively, in Sections 4.1 and 4.2.

### 4.1 Methodology

In this section, we describe the methodology on how we derive the consequences and fix patterns.

#### 4.1.1 Deriving Consequences

We analyze commit messages and content in issue reports for the identified 719 defects to determine the consequences using open coding [68]. The first and last author conducts open coding separately. Each rater applies the following steps: (i) separate commits/issues labeled as defects identified from Section 3.1.2; (ii) read text in commits/issues; (iii) separate text that expresses consequences; (iv) categorize consequences based on commonality, e.g., two issues reports [3] express an outage–related consequence.

Initially, the first and last author respectively, identifies 17 and 12 consequences. As the approach is susceptible to rater bias, we use the last author of the paper for rater verification. The last author performs closed coding [68] where they map the defects to the derived consequences. The Cohen's

---

1. https://github.com/SeldonIO/seldon–core/issues/3677
2. https://github.com/apache/openwhisk–deploy–kube/commit/720abadb5249eb96d5f27afd1cc21387ab85652d
3. https://github.com/argoproj/argo–cd/issues/10249, https://github.com/Azure/application–gateway–kubernetes–ingress/issues/67

Kappa [17] is 0.53, suggesting a 'moderate' agreement [51]. The disagreement resulted from the last author's opinion of merging 5 consequences with the remaining 12. Upon discussion, both raters agree to merge these 5 consequences and end up with the final set of 12 consequences. Each identified defect category from Section 3.2 maps to one of the 12 consequences. These consequences show how serious Kubernetes configuration defects are and highlight the importance of our study.

TABLE 3: Answer to RQ1: Frequency of Defect Categories.

| Category | Sub-category | Count |
|---|---|---|
| Conditional | N/A | 40 |
| Container Provisioning | CLA | 43 |
| | Resources | 9 |
| | Category Total | 52 |
| Custom Resource | N/A | 46 |
| Data Fields | BSE | 2 |
| | IDT | 19 |
| | IUPT | 1 |
| | Syntax | 35 |
| | VR | 30 |
| | Category Total | 87 |
| Entity Referencing | N/A | 125 |
| Incorrect Helming | N/A | 13 |
| Namespaces | N/A | 15 |
| Orphanism | N/A | 10 |
| Pod Scheduling | N/A | 12 |
| Probing | N/A | 22 |
| Property Annotation | N/A | 12 |
| Security | AC | 76 |
| | ESD | 4 |
| | PP | 1 |
| | SC | 11 |
| | Category Total | 92 |
| Unsatisfied Dependency | N/A | 105 |
| Version Incompatibility | N/A | 58 |
| Volume Mounting | N/A | 30 |

#### 4.1.2 Deriving Fix Patterns

We apply a qualitative analysis technique called open coding [68] similar to prior research [42], [92]. The first and last author individually applies the following steps: (i) separate issues labeled as defects from Section 3.1 that have code changes; (ii) read the code that was changed for each defect; (iii) identify commonalities in the changes and create groups based on commonalities; and (iv) merge groups into fix pattern categories. Each rater uses messages and code changes from commits as well as from issue reports to apply the above-mentioned steps.

Upon applying open coding, the first and last author respectively, identifies 12 and 8 fix pattern categories. The authors disagree on 3 categories. Upon discussion, the 8 categories identified by both raters and one category identified by the first author that was not identified by the last authors was added. The Cohen's Kappa [17] is 0.81, suggesting an 'substantial' agreement [51].

### 4.2 Answer to RQ2

We provide answers to RQ2 in this section.

#### 4.2.1 Answer to RQ2: Consequences

We identify 12 consequences, definitions of which are provided in Table 4. A mapping between the identified defect categories and the consequences is provided in Table 5. We observe the most frequently occurring consequence to be incorrect operations (InOp). We observe 52 defects to be related with configuration inexecutability that does not lead to crashes and hangs but keep the Kubernetes cluster running with incorrect configurations.

TABLE 4: Results for RQ2: Consequences and their definitions.

| Consequence | Definition |
|---|---|
| Compiler Warning (CW) | The consequence of obtaining warning messages from the compilation engine. |
| Configuration Inexecutability (CI) | The consequence of running the Kubernetes cluster with incorrect configurations. In this case, configurations specified in scripts are not executed or are overridden. |
| Crash | The consequence of a Kubernetes operation being terminated abruptly. |
| Diagnose Inability (DI) | The consequence of not being able to diagnose failures or crashes. |
| Exposure of Unauthorized Data (EUD) | The consequence when unauthorized users get access to data. |
| Hang | The consequence when an operation is unresponsive. |
| Incorrect Artifact Generation (IAG) | The consequence of generating an artifact incorrectly because of a defect. |
| Incorrect Operations (InOp) | The consequence when Kubernetes-related operations are executed incorrectly. |
| Incorrect Rendering (IR) | The consequence of generating an incorrect display for the Kubernetes dashboard. |
| Outage | The consequence when a Kubernetes object is unavailable when requested by users. |
| Performance | The consequence of incurring unexpected usage of CPU and memory. |
| Unpredictable Responses (UR) | The consequence of providing unpredictable responses to the user, such as conducting unpredictable routing of traffic and obtaining unpredictable responses from pods. |

#### 4.2.2 Results for RQ2: Fix Patterns

We identify 9 fix patterns, definitions of which are provided in Table 6. A mapping between the identified defect categories and the fix patterns is provided in Table 7. The most frequently occurring fix pattern to be configuration value changes (CVC).

## 5 RQ3: EVALUATION OF STATIC ANALYSIS TOOLS FOR DETECTING DEFECTS

We organize this section by answering two sub-questions:

- RQ3.a: What categories of Kubernetes-related defects are supported by static analysis tools?

- RQ3.b: How can we detect defects that are not supported by existing static analysis tools?

We evaluate static analysis tools in our paper. We do not evaluate dynamic analysis tools, such as 'Kube-hunter' [4] and 'BotKube' [5] as these tools rely on logs that are generated from a Kubernetes cluster at runtime. Therefore, evaluation of these dynamic analysis tools require execution of configuration scripts, which in turn is dependent on correct inference of computing environments [57]. Setting up these environments correctly require adequate installation of all artifacts specified as dependencies for each of the 185 repositories, which makes the evaluation of dynamic analysis tools unfeasible.

---

4. https://github.com/aquasecurity/kube-hunter
5. https://botkube.io/

TABLE 5: Answer to RQ2: Frequency of consequences. '–' means zero defects map to that consequence.

| Defect Category | CW | CI | Crash | DI | EUD | Hang | IAG | InOp | IR | Outage | Performance | UR |
|---|---|---|---|---|---|---|---|---|---|---|---|---|
| Conditional | – | 9 | 13 | – | – | 1 | 9 | 1 | – | 6 | – | 1 |
| Container Provisioning | – | 7 | 4 | 1 | – | 2 | 1 | 8 | – | 24 | 4 | 1 |
| Custom Resource | – | 3 | 13 | 6 | – | 1 | 2 | 6 | – | 13 | 2 | – |
| Data Fields | 1 | 3 | 57 | – | – | 3 | – | 6 | – | 15 | 2 | – |
| Entity Referencing | – | 19 | 34 | 5 | – | 1 | 2 | 25 | 2 | 26 | 7 | 4 |
| Incorrect Helming | – | 8 | 1 | – | – | – | – | – | – | 3 | – | 1 |
| Namespaces | – | – | 1 | – | – | – | – | 11 | – | 3 | – | – |
| Orphanism | – | – | 1 | – | – | – | – | 2 | – | – | 7 | – |
| Pod Scheduling | – | 1 | 1 | – | – | – | – | 3 | – | 4 | 3 | – |
| Probing | – | 1 | 3 | – | – | – | – | – | – | 12 | 2 | 4 |
| Property Annotation | – | – | 2 | 2 | – | 1 | – | 3 | – | 3 | 1 | – |
| Security | 1 | – | 2 | 4 | 9 | 1 | – | 63 | – | 12 | – | – |
| Unsatisfied Dependency | – | 1 | 6 | 7 | – | 6 | – | 46 | – | 29 | 7 | 3 |
| Version Incompatibility | 7 | – | 17 | 2 | – | 1 | 1 | 13 | – | 15 | – | 2 |
| Volume Mounting | 1 | – | 6 | 2 | – | 1 | – | 7 | – | 13 | – | – |
| Total | 10 | 52 | 161 | 29 | 9 | 18 | 15 | 194 | 2 | 178 | 35 | 16 |

TABLE 6: Answer to RQ2: Fix patterns and their definitions.

| Fix Pattern | Definition |
|---|---|
| Adding Conditional Statements (ACS) | This fix pattern corresponds to adding conditional statements. |
| Configuration Value Changes (CVC) | This fix pattern corresponds to changing configuration values. |
| Directive Fix (DF) | This fix pattern corresponds to fixing a template directive in order to populate an YAML file with the correct configuration values. |
| Environment Variable Fix (EVF) | This fix pattern corresponds to changing environment variables used for the container runtime. |
| Object Modification (OM) | This fix pattern corresponds to the creation or deletion of Kubernetes objects. |
| Property Modification (PM) | This fix pattern corresponds to property addition or property deletion for a certain Kubernetes object. |
| Relocation | This fix pattern corresponds to relocation of Kubernetes objects, paths, and properties. |
| Rule Fix (RF) | This fix pattern corresponds to rules used for setting up access control policies using apiGroups, resources, and/or verbs, such as 'get', 'list', 'create', and 'delete'. |
| Syntax Fix (SF) | This fix pattern corresponds to fixing syntax issues. |

In order to conduct evaluation, we use the curated dataset described in Section 3.1. This dataset is informed by: (i) real-worlds defects confirmed by practitioners; and (ii) manual verification by the raters. An alert reported by a tool that does not exist in the dataset is a false positive. Any defect included in the dataset but missed by the tool is a false negative. Our approach is consistent with prior research [52] that conducted tool evaluation using curated datasets.

## 5.1 RQ3.a: Defect Categories Supported by Static Analysis Tools

### 5.1.1 Methodology

We use the following steps to answer RQ3.a:

5.1.1.1 Selection of Static Analysis Tools: We start the selection process using the Google search engine in incognito mode with the search string 'defect detection tools for kubernetes'. From the collected top 100 search results, we identify 100 tools for Kubernetes. Next, we apply the following criteria: Criterion-1: The tool must be publicly available for use. Criterion-2: The tool must be able to detect defects using static analysis. The first author of the paper read the documentation of each tool to determine if the tool can detect defects in configuration scripts. Criterion-3: The tool must support execution through the command line interface, allowing for automated execution. Criterion-4: The tool must be capable of detecting at least one of the 15 identified defect categories. This ensures that each tool contributes to the overall coverage of defect detection. The first author reads the documentation of each tool to apply this criterion. By applying Criterion–1, 2, and 3, we respectively, identify 23, 20, and 8 tools. From our application of the four criteria we identify eight tools. Attributes of these tools are available in Table 8. Each of the 8 tools were applied on 2,260 scripts using the command line. For example, 'Kubeconform' was executed using 'kubeconform < file_path >'. The process took 9.75 hours in total, averaging 1.2 hours per tool.

5.1.1.2 Evaluation of Static Analysis Tools: We use two evaluation activities:

**Activity-1: Evaluation based on support**: For this evaluation, we conduct a mapping between each identified defect categories to a detection rule used by each of the eight tools. The second author applies closed coding [68] where they read the documentation and source code of each tool to perform this mapping. A mapping exists if a rule matches the definition of a defect category.

**Activity-2: Evaluation based on detection accuracy**: Using precision and recall, we compute the detection accuracy of the generated alerts, i.e., the detection results obtained from each tool. Precision is calculated as $\frac{TP}{TP+FP}$. Recall is calculated as $\frac{TP}{TP+FN}$. Here, $TP$ corresponds to a true positive alert, whereas FP corresponds to a false positive alert. We determine an alert to be a TP if the alert correctly identifies a defect for the same category, same configuration script, same location, and same coding pattern for the defect of interest. We determine an alert to be a FP if the alert incorrectly identifies a defect belonging to an incorrect category, or incorrect script, or incorrect location, or for the incorrect coding pattern. We determine FN for a defect in our dataset used for RQ1, if a tool does not report an alert for it. In order to determine TP, FP, and FN we use the dataset that we construct for answering RQ1. We do not any include defect that is not present in our dataset used for categorization. We repeat the process for calculating TP, FP, and FN for all defect categories.

TABLE 7: Answer to RQ2: Frequency of fix patterns. '–' means no defects map to the fix pattern.

| Defect Category | ACS | CVC | DF | EVF | OM | PM | Relocation | RF | SF |
|---|---|---|---|---|---|---|---|---|---|
| Conditional | 13 | – | 27 | – | – | – | – | – | – |
| Container Provisioning | 1 | 32 | 3 | 9 | 1 | 4 | 1 | 1 | – |
| Custom Resource | 3 | 9 | 3 | – | 2 | 29 | – | – | – |
| Data Fields | 4 | 8 | 23 | 1 | 4 | 12 | 1 | – | 34 |
| Entity Referencing | 10 | 58 | 41 | 2 | – | 7 | 4 | 3 | – |
| Incorrect Helming | – | – | 9 | 1 | – | 3 | – | – | – |
| Namespaces | – | 2 | 3 | 1 | 1 | 6 | – | 2 | – |
| Orphanism | – | 3 | 1 | – | 5 | – | – | 1 | – |
| Pod Scheduling | 2 | – | 1 | – | – | 9 | – | – | – |
| Probing | 1 | 6 | – | – | – | 15 | – | – | – |
| Property Annotation | 2 | 6 | 1 | – | – | 3 | – | – | – |
| Security | 1 | 4 | 3 | – | 2 | 11 | – | 71 | – |
| Unsatisfied Dependency | 11 | 11 | 2 | 5 | 16 | 25 | 2 | 33 | – |
| Version Incompatibility | 14 | 32 | 5 | – | 1 | 4 | 2 | – | – |
| Volume Mounting | 4 | 3 | 1 | – | – | 20 | 2 | – | – |
| Total | 66 | 174 | 123 | 19 | 32 | 148 | 12 | 111 | 34 |

TABLE 8: Descriptions and Attributes of Selected Static Analysis Tools

| Tool | Description | Source | Output Format | Size(LOC) |
|---|---|---|---|---|
| Checkov | A tool that can scan configurations used in cloud infrastructure. It supports over 1,000 checks related to security and compliance. | GitHub [13] | SARIF, Text, JSON, XML, CSV, Markdown | 695,709 |
| Datree | A tool designed to secure Kubernetes workloads. It focuses on workload security, resource management, and best practices. | GitHub [22] | SARIF, JSON, XML, Text | 31,193 |
| Kube-Score | A tool designed to analyze Kubernetes object definitions. It investigates Kubernetes resources and provides recommendations to enhance the resilience of applications. | GitHub [89] | SARIF, JSON, JUnit, Text, CI | 17,054 |
| KubeLinter | A tool developed that identifies security defects and deviations from recommended practices. KubeLinter is perceived as the most popular static security analysis tool. | GitHub [75] | SARIF, JSON, Text | 24,295 |
| Kubesec | A security–focused static analysis tool that identifies potential security weaknesses in configuration scripts. It assigns a security score to Kubernetes resources based on their configuration. | GitHub [18] | JSON, YAML, Text | 9,919 |
| Kube-conform | A tool that validates scripts with OpenAPI and JSON schemas, ensuring they comply with expected standards. | GitHub [86] | JSON, XML, Text, TAP | 639,910 |
| SLI-KUBE | A tool developed by researchers that identifies 11 categories of security weaknesses in scripts. It can be executed from the command line and is available as a Docker image. | TOSEM'23 [64] | SARIF, CSV | 10,987 |
| Yamlint | A tool that checks for syntax validity and adherence to best practices, including key repetition and syntax issues, such as trailing spaces. | GitHub [1] | Text | 11,535 |

### 5.1.2 Results for RQ3.a

Our results are:

**5.1.2.1 Results Related to Support**: We find eight defect categories to be supported by at least one tool. The defect categories for which we observe no support are: conditional, CR, incorrect Helming, orphanism, property annotation, unsatisfied dependency, and volume mounting. A full breakdown is available in Table 9, which is organized alphabetically by category names. In the table, a '✓' indicates that the tool can detect the category, while a '–' denotes that the tool cannot detect the category.

**5.1.2.2 Results related to Detection Accuracy**: We observe the average precision and recall to be ≤ 0.28 for all eight tools. The highest precision is observed for Datree and Kubesec respectively, for syntax and incorrect data types (IDT), which are sub–categories of data fields. The highest recall is observed for Yamllint in the case of detecting defects–related to syntax. The worst performing tool is SLI–KUBE as its precision and recall is 0.0 for all categories. Data related to all tools and defect categories are available in Table 10. The '#' column represents the count of defects for each category and sub–category.

### 5.2 RQ3.b: Defect Detection with ConShifu

We provide the methodology and results for RQ3.b respectively, in Sections 5.2.1 and 5.2.2.

### 5.2.1 Methodology

Answers to RQ3.a show that there are seven categories of defects that are not covered by any tool. Of these seven categories, incorrect Helming and orphanism can be detected using static analysis. Detection of these two categories

TABLE 9: Answer to RQ3.a: Support for Detecting Defects in Kubernetes Configuration Management.

| Category | Sub-category | Checkov | Datree | Kube-conform | Kube-Linter | Kube-Score | Kubesec | SLI-KUBE | Yaml-lint |
|---|---|---|---|---|---|---|---|---|---|
| Conditional | N/A | – | – | – | – | – | – | – | – |
| Container Provisioning | CLA | – | – | – | – | – | – | – | – |
| | Resources | ✓ | ✓ | – | ✓ | ✓ | ✓ | ✓ | – |
| Custom Resource | N/A | – | – | – | – | – | – | – | – |
| Data Fields | BSE | – | – | – | – | – | – | – | – |
| | IDT | – | ✓ | ✓ | – | – | ✓ | – | – |
| | IUPT | – | – | – | – | – | – | – | – |
| | Syntax | – | – | – | – | – | – | – | ✓ |
| | VR | – | ✓ | ✓ | ✓ | ✓ | ✓ | – | – |
| Entity Referencing | N/A | ✓ | ✓ | – | ✓ | ✓ | ✓ | – | – |
| Incorrect Helming | N/A | – | – | – | – | – | – | – | – |
| Namespaces | N/A | ✓ | ✓ | – | ✓ | – | – | ✓ | – |
| Orphanism | N/A | – | – | – | – | – | – | – | – |
| Pod Scheduling | N/A | – | ✓ | – | ✓ | ✓ | – | – | – |
| Probing | N/A | ✓ | ✓ | – | ✓ | ✓ | – | – | – |
| Property Annotation | N/A | – | – | – | – | – | – | – | – |
| Security | AC | ✓ | ✓ | – | ✓ | – | ✓ | – | – |
| | ESD | ✓ | ✓ | – | ✓ | – | – | ✓ | – |
| | PP | – | – | – | – | – | – | – | – |
| | SC | ✓ | ✓ | – | ✓ | ✓ | ✓ | ✓ | – |
| Unsatisfied Dependency | N/A | – | – | – | – | – | – | – | – |
| Version Incompatibility | N/A | – | ✓ | – | ✓ | ✓ | – | – | – |
| Volume Mounting | N/A | – | – | – | – | – | – | – | – |

TABLE 10: Detection accuracy of eight tools. '–' means a precision (P) or recall (R) of 0.0.

| Category | Sub-category | # | Checkov | | Datree | | Kube-conform | | Kube-Linter | | Kube-Score | | Kubesec | | SLI-KUBE | | Yaml-lint | |
|---|---|---|---|---|---|---|---|---|---|---|---|---|---|---|---|---|---|---|
| | | | P | R | P | R | P | R | P | R | P | R | P | R | P | R | P | R |
| Container Provisioning | Resources | 9 | 0.01 | 0.11 | – | – | – | – | 0.01 | 0.12 | 0.002 | 0.05 | 0.02 | 0.27 | – | – | – | – |
| Data Fields | IDT | 19 | – | – | 0.02 | 0.03 | 0.67 | 0.03 | – | – | – | – | 1.00 | 0.03 | – | – | – | – |
| | Syntax | 35 | – | – | 1.00 | 0.01 | – | – | – | – | – | – | – | – | – | – | 0.001 | 0.50 |
| | VR | 30 | – | – | 0.24 | 0.09 | 0.24 | 0.07 | – | – | 0.24 | 0.07 | 0.28 | 0.07 | – | – | – | – |
| Entity Referencing | N/A | 125 | 0.002 | 0.003 | – | – | – | – | 0.01 | 0.01 | 0.03 | 0.01 | 0.01 | 0.01 | – | – | – | – |
| Namespaces | N/A | 15 | 0.01 | 0.14 | – | – | – | – | – | – | – | – | – | – | – | – | – | – |
| Probing | N/A | 22 | 0.06 | 0.20 | 0.03 | 0.06 | – | – | – | – | 0.06 | 0.16 | – | – | – | – | – | – |
| Security | AC | 76 | 0.02 | 0.02 | – | – | – | – | – | – | – | – | – | – | – | – | – | – |
| | ESD | 4 | 0.002 | 0.17 | – | – | – | – | – | – | – | – | – | – | – | – | – | – |
| | SC | 11 | 0.01 | 0.20 | – | – | – | – | – | – | 0.01 | 0.13 | 0.02 | 0.39 | – | – | – | – |
| Version Incompatibility | N/A | 58 | – | – | – | – | – | – | 0.04 | 0.02 | 0.08 | 0.01 | – | – | – | – | – | – |
| **Avg.** | | 404 | 0.01 | 0.01 | 0.02 | 0.01 | 0.28 | 0.004 | 0.01 | 0.01 | 0.02 | 0.01 | 0.02 | 0.01 | – | – | 0.001 | 0.001 |

of defects is important as these defects can cause crashes and outages, as shown in Table 5. We hypothesize that by leveraging coding patterns from existing defects related to these two categories, we can develop a linter for defect detection. Accordingly, we construct 'ConShifu' [6] using the following steps:

**Step#1 - Parsing**: ConShifu takes one or multiple configuration scripts as input. Each script is parsed into key–value pairs where the hierarchies of keys are preserved. ConShifu is capable of analyzing Kind and Helm scripts. Upon completion of parsing, ConShifu stores the output in the forms of key–value pairs in JSON files.

**Step#2 - Rule Matching**: After parsing is complete, ConShifu applies rule matching to identify defects similar to existing static analysis tools [66]. The rules are listed in Table 11. String patterns needed to implement 'isKind' is shown in the 'String Pattern' column. For rule derivation, we identify commonalities amongst coding patterns that map to existing defects reported in Section 3.1. For example, the coding patterns mountPath: /var/lib/kubelet and mountPath: /var/lib/kubelet/plugins/ebs.csi.aws.com appear for two instances of incorrect Helming where a hard–coded value is used for a key called 'mountPath'. The commonality here is both coding patterns having a hard–coded value for a key that is used in a template. Thus, we can abstract these coding patterns into a rule 'isTemplate(x) ∧ ∃((x.key) ∧ isHardCoded(x.key.value))'. We repeat the same process for orphanism.

---

6. 'Shifu' (师傅) is a Chinese word, which means 'master'

Con–Shifu is a Python–based tool that we execute using the command line for 2,260 scripts in 0.76 hours. Using ConShifu we identify 381 instances of defects. We use a random sample with 95% confidence from the set of 381 instances. For the selected set of 192 instances we obtain an average precision and recall of respectively, 0.83 and 0.92. The precision and recall of ConShifu for incorrect Helming is respectively, 0.85 and 0.96. The precision and recall of ConShifu for orphanism is respectively, 0.81 and 0.89. These results provide us the confidence that the detected instances of incorrect Helming and orphanism could be of relevance to practitioners.

**Step#3 - Evaluation Using Practitioner Feedback**: We submit issue reports to obtain feedback on the detected defects by ConShifu. We apply ConShifu on 124 repositories that are active as of August 01, 2024. ConShifu analyzes 8,576 scripts in 22 minutes and respectively, identifies 183 and 198 instances of incorrect Helming and orphanism. We take a random sample and submit 24 issue reports for 26 instances of incorrect Helming and 18 instances of orphanism. We take a random sample to comply with ethical recommendations by not spamming the practitioners [29]. Each issue report includes the defect's location, a brief description, the potential consequences, and a fix that is submitted as a GitHub pull request.

TABLE 11: Rules Used by ConShifu

| Category | Rule | String Pattern |
|---|---|---|
| Incorrect Helming | isTemplate(x) ∧ ∃((x.key) ∧ isHardCoded(x.key.value)) | N/A |
| Orphanism | (isKind(x) ∧ ¬isReferenced(x.key.value)) ∨ (isKind(x) ∧ ¬isReferenceExist(x.key.value)) | 'ServiceAccount', 'ClusterRole', 'StorageClass', 'PersistentVolumeClaim', 'PersistentVolume', 'Role' |

*5.2.2 Results for RQ3.b*

As of Jan 20 2025, we obtain 33 responses for 44 defects. Practitioners have confirmed 26 defects as valid. As shown in Figure 6, of the 26 valid defects, 21 are related to incorrect Helming and 5 are related to orphanism. Four defects of orphanism detected by ConShifu are rejected as they reside in an application where the configuration values are expected to be provided by users. Evidence of submitted defect reports are available online [91].

## 6 DISCUSSION

We discuss the implications and limitations respectively, in Sections 6.1 and 6.2.

### 6.1 Implications of Our Findings

The implications of our findings are:

6.1.0.1 **'Shift Left' Approach Towards Defect Detection**: In software development, the 'shift left' approach advocates for pro–active integration of quality assurance activities, such as application of static analysis tools in the software development process [60]. We advocate for a 'shift left' approach for configuration management of Kubernetes as well. From our analysis, we observe 533 of the 719 defects result in a crash or an incorrect operation or outage. This finding shows defects in Kubernetes configuration scripts to be consequential, and therefore the community should take actions on how to facilitate defect detection for Kubernetes configuration scripts. Our findings and the dataset could be helpful in this regard as it could help the community understand the nature of defects. While static analysis tools suffer from low actionability due to false positives [67], these tools still provide value for practitioners [3], and therefore could be useful for detecting configuration defects.

6.1.0.2 **The Need for Enhancing Static Analysis Tools for Kubernetes**: According to our analysis, none of the studied tools have support for 7 of the 15 categories. We also observe the second most frequently occurring defect category is unsatisfied dependency for which none of the eight studied tools provide any support. With a precision value of 0.28, Kubeconform has the highest average precision amongst all 8 tools. This is lower than what practitioners perceive 'acceptable', i.e., a precision <= 0.90 [67]. Furthermore, while 5 of the 8 studied tools support the most frequently occurring category of entity referencing, the precision and recall is ≤ 0.03 for each tool.

The above–mentioned evidence highlights the need of enhancing static analysis tools for Kubernetes with respect to support and increasing detection accuracy. We provide three recommendations. **First**, detection rules used by existing tools need to be improved. Our curated dataset of defects can be used for improving the rules. **Second**, practitioner feedback can be collected to improve the detection accuracy of static analysis tools. These tools should allow for seamless integration into existing developer workspace in order to collect feedback for the detected defects. Prior research also advocated for obtaining practitioner feedback to improve detection accuracy of static analysis tools [66]. **Third**, runtime data from Kubernetes clusters can be collected to detect five categories of defects namely, conditional, CR, property annotation, unsatisfied dependency, and volume mounting. Detection for each of these categories is dependent on information that can be collected at runtime. An example utility is 'kubectl cluster–info' that can provide cluster information at runtime [47].

6.1.0.3 **Implications for Defect Repair**: Our findings highlight the need of developing automated defect repair tools for Kubernetes configuration scripts. The top four most frequently occurring fix patterns are configuration value changes, directive fix, property modification, and rule fix that are applied manually to fix 553 out of 719 defects. In order to develop defect repair techniques, researchers can use the curated list of defects and their corresponding fixes that are available as part of our dataset. We posit prior automated defect repair techniques to under–perform for 8 of the 15 defect categories that have not been reported in prior software systems.

6.1.0.4 **Implications of Defects Related to Configuration Inexecutability**: From Section 4.2.1, we observe 52 defects to be related with configuration inexecutability. We find these defects to not exhibit any explicit symptoms, such as crashes or outages, which makes the defect detection process challenging. According to our analysis, practitioners take a reactive approach where they use the 'kubectl' command manually to identify these defects. This approach is time consuming, which necessitates development of automated techniques. One possible future direction can be usage of existing log–based defect localization techniques [21]. Another possible future direction could be application of reachability analysis [90] to detect defects related to configuration inexecutability.

### 6.2 Threats to Validity

We discuss the limitations of our paper as follows:

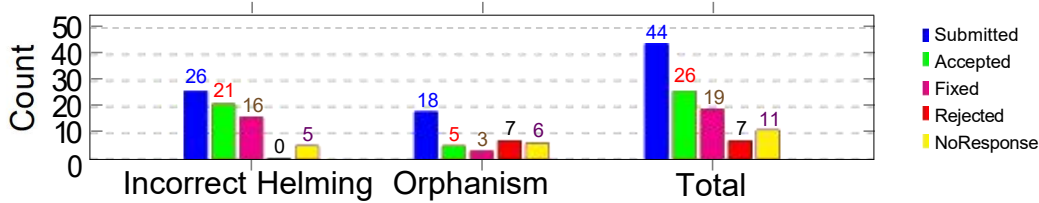

Fig. 6: Count of submitted, accepted, and fixed defects identified by ConShifu.

Conclusion Validity: Answers to RQ1, RQ2, and RQ3 are subject to rater bias as the first author of the paper applies necessary qualitative analysis. We mitigate this limitation by performing rater verification. Answers to RQ3 is limiting, as we use eight tools and may have missed tools not included our paper. We mitigate this limitation by using a systematic selection criteria. Furthermore, evaluation results for studied tools is dependent on the dataset created in Section 3.1, which many bias the results.

Construct Validity: Our study is susceptible to construct validity as the defect identification process depends on the accuracy and completeness of the parsed scripts. ConShifu is susceptible to miss defects as it uses a rule–based approach to identify defects. Furthermore, ConShifu can generate false positives while reporting instances of incorrect Helm-ing and orphanism. ConShifu may fail to detect instances of incorrect Helm if there are no Helm scripts, values .yaml files, or templates.

External Validity: Our findings are obtained from OSS repositories, which may not generalize for configuration scripts used in proprietary repositories. We mitigate this limitation by analyzing repositories from GitHub, which is the most popular code sharing platform.

## 7 RELATED WORK

Our paper is related with existing research on defect categorization and quality assurance aspects of Kubernetes, which we describe in the following subsections:

### 7.1 Prior Research Related with Defect Categorization

Software defect categorization has been of interest to researchers since the 1990s. In 1992, Chillarege and colleagues [15] proposed the orthogonal defect classification (ODC) taxonomy, which consists of eight defect categories. Since then, researchers have used and extended the ODC taxonomy. For instance, Alannsary and Tian [2] and Silva et al. [73] used ODC to respectively, categorize defects for software–as–a–service and embedded software systems. ODC was also extended by Hunny et al. [36] to classify security vulnerabilities.

Researchers have also developed their own taxonomies because of ODC's limitations for modern software systems [72]. Researchers, such as Yu et al. [87], Wan et al. [78], Cui et al. [20], and Du et al. [25] in separate publications derived defect taxonomies respectively, for container runtime systems, blockchain projects, database systems, and federated learning systems. Makhshari and Mesbah [56], Chen et al. [14], Shen et al. [71], Gao et al. [28], Wang et. al [80], Wang et. al [79] constructed defect taxonomies respectively, for IoT software projects, deep learning–based deployment, deep learning compilers, distributed systems, android applications, and autopilot software systems. Wang et al. [81] analyzed 83 defects in WeChat Mini–Programs, and categorized them into 6 categories. Cotroneo et al. [19] categorized the failures of OpenStack using a bottom–up approach. Hassan et al. [34] conducted an empirical study involving 5,110 state reconciliation defects and classify these defects into 8 categories. Rahman et al. [63] developed a taxonomy of defects in IaC scripts by applying descriptive coding with 1,448 defect–related commits. Humbatova et al. [35] analyzed GitHub issues and Stack Overflow posts to develop a classification of faults for software projects involving deep learning. Wang et al. [82] studied configuration defects that occur when these configurations are provided at runtime for database and web server systems.

### 7.2 Prior Research Related with Quality Aspects of Kubernetes

Researchers have shown increasing interest in quality assurance for Kubernetes in recent years. Yang et al. [85] focused on vulnerabilities in the orchestration layer, and recommended two practices for enhancing the security of Kubernetes clusters. Kamieniarz et al. [43] studied the security vulnerabilities that can occur in Kubernetes–related deployments. Rahman et al. [64] in particular identified what types of Kubernetes objects are impacted by security weaknesses, such as hard–coded passwords and insecure HTTP. They [64] also quantified correlations between development activity metrics and the presence of security weaknesses. Carmen et al. [11] in their study, created a new taxonomy for Kubernetes scheduling techniques, organizing the techniques into five main domains and highlighting where current scheduling techniques fall short, especially in terms of security and performance. Gu et al. [32], Sun et al. [76], [77], and Xu et al. [84] in separate publications focused on analyzing and detecting defects related to Kubernetes controllers. Xu et al. [84] focused on deriving a taxonomy for defects that occur in Kubernetes operators, which are specialized controllers. Gu et al. [32] and Sun et al. [76], [77] focused on deriving testing techniques that can expose defects in Kubernetes controllers and operators.

The closest in spirit is prior research that have focused on operator–related defects [32], [76], [77]. Operators extend the functionality of Kubernetes, e.g., the Go–based Spark operator [7] enables the execution of Spark applications on Kubernetes. Our paper focuses solely on defects that occur in YAML–based configuration files, which have no overlap with Kubernetes operators. These configuration files are solely used to provide necessary configurations by Kubernetes users. While Rahman et al. [64] has studied configuration files, their paper only focuses on security–related configuration defects, and did not identify any of the other 14 defect categories reported in our paper. While working with Kubernetes, practitioners need to use configuration options related to (i) pods and (ii) state reconciliation. To configure pods, i.e., abstractions to group containers, practitioners need to understand non–trivial concepts such as affinity and annotations. Likewise, to configure state reconciliation, developers need to understand concepts such as custom resources. Erroneous usage of the these configuration options can result in defective Kubernetes deployments. Compared to prior Kubernetes–related work, we advance knowledge by creating (i) a configuration defect taxonomy, (ii) a novel benchmark; and (iii) a linter that has detected defects confirmed by practitioners. We have identified 7 categories of configuration defects that have not been reported in any prior work: custom resource, incorrect helming, namespaces, orphanism, pod scheduling, property annotation, and volume mounting. Furthermore, we are the first to conduct tool evaluation for identified defect categories.

---

7. https://github.com/kubeflow/spark–operator

## 8 Conclusion

Kubernetes is becoming popular in industry as a tool for automated management of containers. Configuration defects in Kubernetes can be consequential and, unfortunately, are not uncommon. This paper reports an empirical study about Kubernetes–related configuration defects alongside their consequences and fix patterns. The goals of this empirical study are (i) to help practitioners who use Kubernetes to detect configuration defects, and (ii) to offer researchers opportunities for improving existing static analysis tools for detecting those defects. Our study includes 719 defects mined from 185 OSS repositories. We identify 15 defect categories for Kubernetes configuration scripts. We find that insights obtained from existing defects can be used to identify previously–unknown defects. For example, using our linter ConShifu, we identify 26 defects that have been accepted as valid defects by the practitioners of the corresponding OSS projects.

Our research study has produced multiple lessons. For example, we provide recommendations for researchers on how existing defects that are available as part of our dataset, can be leveraged to enhance existing static analysis tools and to develop defect repair techniques for Kubernetes. We also advocate for incorporating practitioner feedback and runtime information to improve existing static analysis tools for Kubernetes configuration scripts.

## Data Availability

The datasets and source code used for the paper are publicly–available online as a replication package [91]. URL of the package: https://figshare.com/s/5c63f862a1abd95f7708.

## Acknowledgments

We thank the PASER group at Auburn University for their valuable feedback. This research was partially funded by the U.S. National Science Foundation (NSF) Award # 2247141, Award # 2312321, and Award # CNS–2026928.

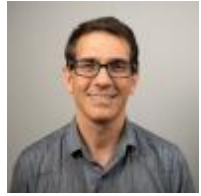

**Marcelo d'Amorim** Marcelo d'Amorim is an Associate Professor in Computer Science at the North Carolina State University, USA. He obtained his PhD from the University of Illinois at Urbana-Champaign in 2007 and his MS and BS degrees from UFPE, Brazil, in 2001 and 1997, respectively. Marcelo's research goal is to help developers build correct software. He is interested in preventing, finding, diagnosing, and repairing software bugs and vulnerabilities.

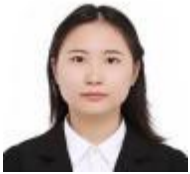

**Yue Zhang** Yue Zhang is a PhD student at Auburn University. Her research interests is in software engineering and data science. She received the B.E. in Computer Science and Technology from the Anhui Jianzhu University, Hefei, China, in 2021.

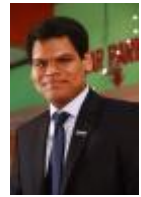

**Akond Rahman** Akond Rahman is an assistant professor at Auburn University. His research interests include DevOps and secure software development. He graduated with a PhD from North Carolina State University, an M.Sc. in Computer Science and Engineering from University of Connecticut, and a B.Sc. in Computer Science and Engineering from Bangladesh University of Engineering and Technology. He won the ACM SIGSOFT Doctoral Symposium Award at ICSE in 2018, the ACM SIGSOFT Distinguished Paper Award at ICSE in 2019, the CSC Distinguished Dissertation Award, and the COE Distinguished Dissertation Award from NC State in 2020. He actively collaborates with industry practitioners from GitHub, WindRiver, and others. To know more about his work visit https://akondrahman.github.io/

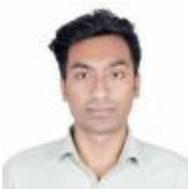

**Uchswas Paul** Uchswas Paul is a Ph.D. student in Computer Science at North Carolina State University, USA. He earned his bachelor's degree from Khulna University of Engineering and Technology in 2018. Before joining his doctorate, he gained experience in industry and academia. His research interests lie in software engineering and large language models.